\documentclass[prb,onecolumn,amssymb,superscriptaddress]{revtex4}
\usepackage{amsmath} 
\usepackage{amssymb} 
\usepackage{amsthm}
\usepackage{graphicx}
\usepackage{caption}
\usepackage{float}
\usepackage{graphicx, subfigure}
\usepackage{epsfig}
\usepackage{epstopdf}
\usepackage{braket}
\usepackage{hyperref}

\usepackage{color}
\usepackage{rotating}
\usepackage{bm}

\begin{document}

\title{Valley Hall phases in Kagome lattices}
\author{Natalia Lera}
\affiliation{Departamento de F\'isica de la Materia Condensada, Universidad Aut\'onoma de Madrid, Madrid 28049, Spain, Condensed Matter Physics Center (IFIMAC) and Instituto Nicol\'as Cabrera}

\author{Daniel Torrent}
\affiliation{GROC, Institut de Noves Tecnologies de la Imatge (INIT), Universitat Jaume I, Castellon 12071, (Spain)}

\author{P. San-Jose}
\affiliation{Materials Science Factory, Instituto de Ciencia de Materiales de Madrid (ICMM-CSIC), Sor Juana In´es de la Cruz 3, 28049 Madrid, Spain}

\author{J. Christensen}
\affiliation{Department of Physics, Universidad Carlos III de Madrid, Leganes 28916, Madrid, Spain}

\author{J.V. Alvarez} 
\affiliation{Departamento de F\'isica de la Materia Condensada, 
Universidad Aut\'onoma de Madrid, Madrid 28049, Spain, Condensed Matter Physics Center (IFIMAC) and Instituto Nicol\'as Cabrera}

\begin{abstract}
We report the finding of the analogous valley Hall effect in phononic systems arising from mirror symmetry breaking, in addition to spatial inversion symmetry breaking. We study topological phases of plates and spring-mass models in Kagome and modified Kagome arrangements. 
By breaking the inversion symmetry it is well known that a defined valley Chern number arises. We also show that effectively, breaking the mirror symmetry leads to the same topological invariant. 
Based on the bulk-edge correspondence principle, protected edge states appear at interfaces between two lattices of different valley Chern numbers.
By means a plane wave expansion method and the multiple scattering theory for periodic and finite systems respectively, we computed the Berry curvature, the band inversion, mode shapes and edge modes in plate systems. We also find that appropriate multi-point excitations in finite system gives rise to propagating waves along a one-way path only.
\end{abstract}
\maketitle

\section{Introduction}

The unusual properties of  fabricated metamaterials originate from their designed patterns and geometry as opposed to their chemical composition. Specifically,  when created with periodic structures,  the study of wave propagation can be treated similar to electrons in periodic potentials\cite{PoissonRatio,Poisson2,photonic_meta,Sound_heat,meta,christensen_vibrant}. In this way, topological properties studied in electronic band structures \cite{Classification} can be transferred to classical metamaterials. Inspired by topological electronic systems, the search for protected modes in classical wave phenomena has been active in recent years in areas such as photonics \cite{Topo_photonic,edges_photonic}, acoustics \cite{Topo_acoustics,kagome_acoustics} and elastic media \cite{Topo_mechanics1,Topo_mechanics2}. The  bulk-boundary correspondence principle has been proved to hold also in these areas by showing how topological protected waves arise at the edge of systems containing topologically inequivalent phases.   
In particular, mechanical metamaterials present several advantages: 1) the flexibility to create patterns and to modify band structures in metamaterials is much richer than in real solids \cite{ccy_experiment}. 2.) In electronic systems topological features are easier to detect when they occur close to the Fermi energy, which is hard to shift and control.  
On the other hand, mechanical systems can be excited  in a wide range of frequencies, and the excitation can be easily tuned to the frequency of the topological mode.   

 We consider mechanical metamaterials with time reversal symmetry, establishing analogy with the quantum valley Hall effect \cite{QVHE_electronic1,QVHE_electronic2,QVHE_electronic3,QVHE_electronic4}. 
This approach has been successfully achieved in spring-mass models and plate topology \cite{QVHE_mechanics1,Ruzzene_graphene,QVHE_mechanics2,QVHE_mechanics3,ccy} 
as well as in photonics \cite{QVHE_photonics1,QVHE_photonics2,QVHE_photonics3,QVHE_photonics4} 
or acoustics \cite{QVHE_sound1,QVHE_sound2,QVHE_sound3,kagome_acoustics} by breaking the spatial inversion symmetry. 
The existence of topological modes have been shown experimentally \cite{QVHE_experiment1,QVHE_experiment2,ccy_experiment}, along with unusual properties in the absence of backscattering \cite{QVHE_experiment3,QVHE_mechanics2}. In continuous systems like plates, wave guiding through edge modes could have applications for mechanically isolating structures or transferring energy and  information through elastic waves. 

In this article we focus on the Kagome lattice, which has a graphene-like structure with degenerate Dirac cones at inequivalent points of the Brillouin Zone. Recent interest in metamaterials based on Kagome arrangement suggest future applications \citep{Johan_antenna,kagome_springs,kagome_acoustics,kagome_riva}. The wide range of crystalline symmetries and the underlying $C_3$ symmetry of this system provides a playground to test mechanical topology as well as distinguishing basic features that are relevant to topological mechanics. 

We study discrete spring-mass models in the linear regime in addition to continuum systems such as plates. The former systems allow analytic computations which capture the essentials of topology in easy models with couplings between few neighbors. In plates long ranged waves need to be taken into account. The understanding of topological modes could lead to relevant engineering applications, in particular, efficient and controlled wave guiding. Plates will be described in the linear regime by Kirchhoff-Love theory. To endow the plate with a crystalline structure, we attach a lattice of resonators on top. Modifications of the unit cell properties might open gaps in the phononic band structure with non-trivial topology.
Remarkably, the methodology used in this paper to describe flexural waves in plates is not based on commercial software but on the Multiple Scattering Theory (MST) developed in Ref. \citep{Dani_graphene}.

The structure of the paper is as follows, in sec \ref{Sec:II} we describe briefly the methodology for studying flexural waves in plates. In section \ref{Sec:III}, we describe the  distorted Kagome lattice, its symmetries and the parameter space used in this paper. In section \ref{Sec:IV}, topology arising from spatial inversion symmetry is deduced from the spring-mass model and explained via plate physics, we employ ribbons to create topological protected edge states and design finite systems with interesting properties, like one-way wave propagation. In section \ref{Sec:V}, we study the effects of mirror symmetry breaking in a  Kagome lattice. In section \ref{Sec:VI}, we conclude this paper.

\section{Plate physics and Methodology}
\label{Sec:II}
In this section, to present the system and derive the notation, we briefly introduce  the classical theory of flexural waves for thin plates and describe the methodology, following the approach taken by Torrent \textit{et al.} \citep{Dani_graphene} and Chaunsali \textit{et al.} \citep{ccy}. We  consider a  thin plate coupled to a lattice of resonators. The equation of motion for the deformation field, $w$ is a fourth order derivative in real space and we look for solutions harmonic in time: $w(\vec{r},t)=w(\vec{r})e^{i\omega t}$. 
\begin{equation}
\left(D\nabla^4 -\omega^2\rho h\right)w(\vec{r})=-\sum_{\vec{R}_\alpha}\kappa_{\vec{R}_\alpha} \left(w(\vec{R}_\alpha)-z(\vec{R}\kappa_\alpha)\right) \delta (\vec{r}-\vec{R}_\alpha)
\label{Eq_plate}
\end{equation}
where $D=\frac{Eh^3}{12(1-\nu^2)}$ is the plate stiffness, $\rho$ is the volume mass density of the plate, $h$ is thickness and the sum runs over all resonator sites $\vec{R}_\alpha$ within the unit cell. Resonators masses and spring constants are respectively $m_\alpha$ and $\kappa_\alpha$ and their displacements are $z(\vec{R}_\alpha)$, in the direction perpendicular to the plate.
The equation for each resonator is,
\begin{equation}
\omega^2 m_\alpha z(\vec{R}_\alpha)=-\kappa_{\vec{R}_\alpha} \left(w(\vec{R}_\alpha)-z(\vec{R}_\alpha)\right)
\label{Eq_resonator}
\end{equation}

\subsection{Plane Wave Expansion}
In the Plane Wave Expansion method (PWE), the lattice is infinite in two dimensions and the displacement field can be written in terms of Bloch waves,
\begin{equation}
w(\vec{r})=\sum_{\vec{G}} W(\vec{G})e^{-i(\vec{G}+\vec{k})\cdot \vec{r}}
\label{Eq_eigenstates}
\end{equation}
where $\vec{G}=n_1\vec{g}_1+n_2\vec{g}_2$ are the reciprocal lattice vectors,  $n_{1,2}$ are integers and $\vec{g}_j$ are the basis of vectors fulfilling $\vec{a}_i\cdot\vec{g}_j=2\pi \delta_{ij}$, with $\vec{a}_i$ being the lattice vectors. The result is either a search for zeros of a complex function as described in Ref. \citep{Dani_graphene} or a generalized eigenvalue problem as described in Ref. \citep{ccy}. For completeness, we highlight some steps of the derivation. 

\textit{Method 1:} Substituting the resonator equation, Eq. \ref{Eq_resonator} into the plate equation, Eq. \ref{Eq_plate}, we get
\begin{equation}
\left(\nabla^4 -\omega^2\frac{\rho h}{D}\right)w(\vec{r})=-\sum_{\vec{R}_\alpha} \frac{m_\alpha}{D}\frac{\omega_\alpha^2\omega^2}{\omega_\alpha^2-\omega^2} w(\vec{R}_\alpha) \delta (\vec{r}-\vec{R}_\alpha)
\label{Eq_resonatorintoplate}
\end{equation}
where $\omega_\alpha^2(\omega)=\kappa_\alpha/m_\alpha$ and $t_\alpha=\frac{m_\alpha}{D}\frac{\omega_\alpha^2\omega^2}{\omega_\alpha^2-\omega^2}$. Due to the system's periodicity we omit the $\vec{R}$ dependence in masses and spring constants. Substituting the Bloch Ansatz Eq. \ref{Eq_eigenstates} into the previous equation, deriving each independent term in the Fourier summation and integrating over the unit cell we obtain,
\begin{equation}
\left(  \left | \vec{k}+\vec{G}\right |^4-\omega^2 \frac{\rho h}{D}\right)W_{\vec{G}} =\sum_{\vec{G'},\alpha} \frac{t_\alpha}{A_c} e^{i(\vec{G'}-\vec{G})\cdot\vec{R}_\alpha}W_{\vec{G'}}
\label{Eq_midderivation}
\end{equation}
where $a$ is the lattice parameter and $A_c$ is the area of the unit cell. We have used the following identities,
\begin{equation}
\begin{matrix}
\int_{UC} e^{-i(\vec{G'}-\vec{G})\cdot\vec{r}} d\vec{r}=A_c\delta(\vec{G'}-\vec{G});& 
\int_{UC} f(\vec{r})\delta(\vec{r}-\vec{R}_\alpha) d\vec{r}=f(\vec{R}_\alpha)
\end{matrix}
\end{equation}

Now, we write the expected solution expanded on a the Fourier basis,
\begin{equation}
W_\beta=\sum_{\vec{G'}}
W_{\vec{G'}} 
e^{i\vec{G'}\cdot\vec{R_\beta}}
\end{equation}
and substitute $W_{\vec{G}}$ from Eq. \ref{Eq_midderivation},
\begin{equation}
W_\beta=\sum_{\vec{G}} \frac{1}{\left | \vec{k}+\vec{G}\right |^4-\omega^2 \frac{\rho h}{D}}\frac{1}{A_c}\sum_\alpha e^{i\vec{G}\cdot(\vec{R}_\alpha-\vec{R}_\beta)} t_\alpha W_\alpha.
\end{equation}
Therefore a set of $N$ equations with $N$ unknowns can be written, where $N$ is the number of resonators per unit cell.
We find solutions of this system as the zeros of the determinant of the following matrix,
\begin{equation}
A_{\alpha\beta}(\vec{k})=\delta_{\alpha,\beta} -\frac{\gamma_\beta\Omega^2a^2}{1-\Omega^2/\Omega_\alpha^2}\sum_{\vec{G}}\frac{e^{-i\vec{G}\cdot(\vec{R}_\alpha-\vec{R}_\beta)} }{\left | \vec{k}+\vec{G}\right |^4a^4-\Omega^2 a^2}
\label{Eq_bandsDani}
\end{equation}
where we have introduced the dimensionless variables $\Omega^2=\omega^2\rho a^2 h/D$ and $\gamma_\alpha=\frac{m_\alpha}{\rho a^2 h}$.
We evaluate for each $\vec{k}$ and deduce its $\Omega(\vec{k})$ solutions. The null space of the $A$ matrix correspond to mode shapes at the resonator points. \\

\textit{Method 2:} 
We substitute Bloch waves from Eq. \ref{Eq_eigenstates} in the plate Eq. \ref{Eq_plate}. Equating for each mode and integrating over the unit cell we get,
\begin{equation}
A_c\left( D\left | \vec{k}+\vec{G}\right |^4-\omega^2 \rho h\right) W_{\vec{G}}=\sum_\alpha \kappa_\alpha \left( z(\vec{R}_\alpha) -\sum_{\vec{G'}} W_{\vec{G'}} e^{-i(\vec{G'}+\vec{k})\cdot\vec{R}_\alpha} \right)e^{i(\vec{G}+\vec{k})\cdot\vec{R}_\alpha}
\end{equation}
Using Bloch's theorem for the resonators we can refer all resonator displacements to the ones of the one unit cell, $z(\vec{R}_\alpha)=z(\vec{R}_{0\alpha})e^{-i\vec{k}\cdot\vec{R}_\alpha}$. We substitute in previous equation,
\begin{equation}
\left( \left | \vec{k}+\vec{G}\right |^4a^4-\Omega^2 \right) W_{\vec{G}}
=\sum_\alpha \gamma_\alpha \Omega_\alpha^2
 e^{i\vec{G}\cdot\vec{R}_\alpha} 
 \left( z(\vec{R}_{0\alpha})-\sum_{\vec{G'}} W_{\vec{G'}} e^{-i\vec{G'}\cdot\vec{R}_\alpha}\right) 
\label{Eq_plate_ccy}
\end{equation}
and in resonator equation Eq. \ref{Eq_resonator},
\begin{equation}
-\Omega^2 z(\vec{R}_{0\alpha})=\Omega_\alpha^2\left( \sum_{\vec{G}} W_{\vec{G}}e^{-i\vec{G}\cdot\vec{R}_\alpha}-z(\vec{R}_{0\alpha})\right)
\label{Eq_resonator_ccy}
\end{equation}
Where we have used the same dimensionless variables $\Omega$ and $\gamma$ than in method 1.
Now Eq. \ref{Eq_plate_ccy} and \ref{Eq_resonator_ccy} are rewritten in matrix form of dimension $N_G+N$ where $N_G$ is the number of reciprocal vectors taken for the computation (calculations in this paper are made with $N_G=49$) and $N$ is the number of resonators per unit cell. 
\begin{equation}
\begin{pmatrix}
P_{11}&P_{12} \\ 
P_{21} &P_{22} 
\end{pmatrix}\begin{pmatrix}
W_{\vec{G}}\\ 
z(\vec{R}_{0,\alpha})
\end{pmatrix}=\Omega^2\begin{pmatrix}
Q_{11}&0\\ 
0 &Q_{22} 
\end{pmatrix}\begin{pmatrix}
W_{\vec{G}}\\ 
z(\vec{R}_{0,\alpha})
\end{pmatrix}
\label{Eq_eigproblem}
\end{equation}
where 
\begin{equation}
\begin{matrix}
P_{11,ij}= a^4 \left |  \vec{k}+\vec{G}_{i}\right |^4 \delta_{i,j}+\sum_\alpha \gamma_\alpha\Omega_\alpha e^{i (\vec{G}_j-\vec{G}_i)\cdot \vec{R}_\alpha}\\
P_{12,i\alpha}=-\gamma_\alpha\Omega_\alpha^2 e^{i\vec{G}_i\cdot \vec{R}_\alpha}=P_{21,\alpha i}^*\\
P_{22,\alpha\beta}=\gamma_\alpha\Omega_\alpha^2\delta_{\alpha,\beta}\\
Q_{11,ij}=\delta_{i,j}\\
Q_{22,\alpha\beta}=\gamma_\alpha \delta_{\alpha,\beta}.
\end{matrix}
\end{equation}
In. Eq (14) we use $i,j$ indices for the $N_G$ reciprocal vectors and $\alpha,\beta$ for the $N$ resonators of the unit cell.

The generalized eigenvalue problem gives us the band structure, $\Omega(\vec{k})$, and the mode shape by substituting $W_{\vec{G}}$ into Eq. \ref{Eq_eigenstates}. \\


\subsection{Edge states in ribbons}
\begin{figure}
\epsfig{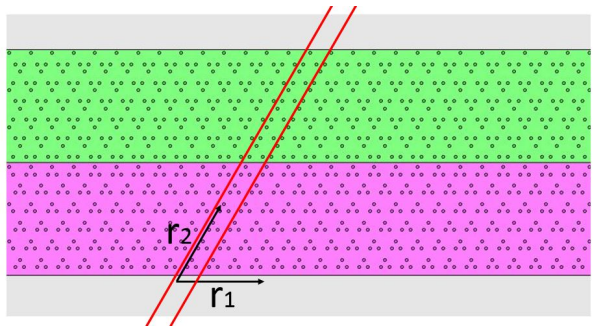}\\
\caption{(Color online) Schematic representation of a ribbon in an infinite plate. $r_1$ and $r_2$ are a basis of lattice. The two red parallel lines delimit one supercell, the supercell is infinite in size. The unit cell is presented with two different topological phases as we will see later in the text.
}
\label{scketch_ribbon}
\end{figure}

We consider ribbons of resonators arranged periodically in the $\vec{r}_1$-direction.  However, the plate is still infinite, so the unit cell in direction $\vec{r}_2$ is infinite, where $\vec{r}_i$ form a basis in 2D. The unit cell is infinite in size but with finite number of resonators present in the supercell, see Fig. \ref{scketch_ribbon}. Unlike electronic systems where wave functions decay exponentially in space, flexural waves decay slowly in the plate and an infinite large unit cell will account for long range waves along the $\vec{r}_2$ direction. The discrete summation over $n_2\vec{g}_2$ in Eq. \ref{Eq_eigenstates} transforms into an integral. 
\begin{equation}
\frac{1}{A_c}
\sum_{G_2}
\rightarrow 
\frac{1}{2\pi a}
\int_{-\infty}^\infty
 dg_2
\end{equation}
Applying this transformation to Eq. \ref{Eq_bandsDani}, $A_{\alpha,\beta}$ matrix simplifies to depend only on $k_1$.
The governing equations are described in Ref. \citep{Dani_graphene}. 
Our main interest creating ribbons consist of studying boundary states between two phases. The interface is contained in the supercell. Bands are computed from the zeros of the $A(\vec{k})$ matrix determinant and its null space contains the eigenmodes, i.e. the $w(\vec{R}_\alpha)$ weight over the supercell resonators. 

\subsection{Multiple Scattering Method}
For finite clusters in an infinite plate we use Multiple Scattering Theory (MST). The governing equations are Eq. \ref{Eq_plate}-\ref{Eq_resonator} where the number of $\vec{R}_\alpha$ is finite. The Green's function of the plate equation without resonators, $G_0(\vec{r})$, is used as a basis to expand the solution of the resulting wave. A system of self-consistent equations lead to the solution of the field $w(\vec{r})$ under some harmonic incident field $\psi_0(\vec{r},t)=\psi_0(\vec{r})e^{i\omega t+\varphi}$
\begin{equation}
w(\vec{r})=\psi_0(\vec{r})+
\sum_\alpha 
T_\alpha
\psi_e(\vec{R}_\alpha) 
G_0(\vec{r}-\vec{R}_\alpha)
\label{Eq_MST1}
\end{equation}
$\psi_e$ is the incident field at scatterer $\alpha$ which allows to deduce the value of $T_\alpha=\frac{t_\alpha}{1-it_\alpha/(8k^2)}$. $\psi_e(\vec{R}_\alpha)$ can be solved from the system of equations,
\begin{equation}
\psi_e(\vec{R}_\alpha)=\psi_0(\vec{R}_\alpha)+\sum_\beta (1-\delta_{\alpha,\beta})T_\beta G_0(\vec{R}_\alpha-\vec{R}_\beta)\psi_e(\vec{R}_\beta)
\label{Eq_MST2}
\end{equation}
We compute the resulting field $w(\vec{r})$ by substituting the solution of 
$\psi_e$ back into Eq. \ref{Eq_MST1}. The incident field is the external excitation of the system and is taken as a point source $\psi_0(\vec{R}_\alpha)=G_0(\vec{R}_\alpha-\vec{x}_0)$, we also consider multipoint dephased excitations $\psi_0(\vec{R}_\alpha)=\sum_j G_0(\vec{R}_\alpha-\vec{x}_j)e^{i\varphi_j}$ and solutions without input field $\psi_0(\vec{R}_\alpha)=0$ that we call natural excitations of the system.

\section{Kagome lattice, distortions and symmetries}
\label{Sec:III}
\begin{figure}
\epsfig{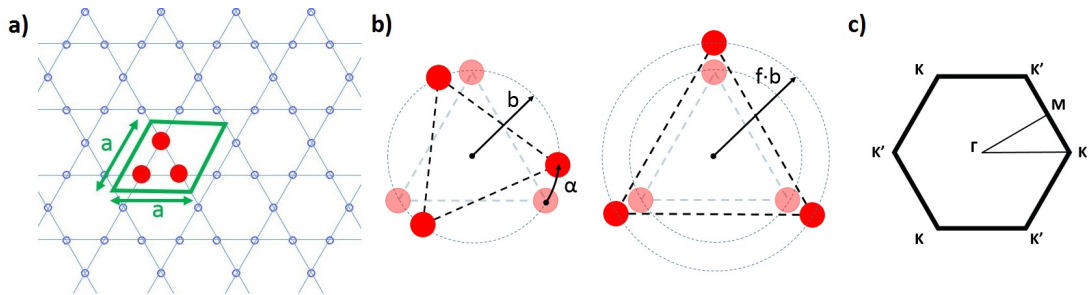}
\caption{(Color online) a) Undistorted Kagome lattice. The unit cell is indicated in a green box of side $a$. The unit cell contains three resonators forming equilateral triangles. b) parameters used in the paper for deformations of Kagome lattice and its effect in the unit cell. They are characterized by an angle $\alpha$ and a uniform expansion factor $f$. c) Brillouin Zone
 }
\label{kagome_lattice}
\end{figure}

The standard Kagome lattice consists of three sets of straight
parallel lines intersecting at lattice sites as shown in Fig. \ref{kagome_lattice}. This figure also shows the unit cell chosen in this article as a parallelogram with lattice vectors
\begin{equation}
\begin{matrix}
\vec{a}_1=a(1,0) & \vec{a}_2=a(\cos(\frac{\pi}{3}),\sin(\frac{\pi}{3})) 
\end{matrix}
\end{equation} 
The normalized masses and resonator frequencies are $\gamma_\alpha=10$ and $\Omega_\alpha=4\pi$ respectively for the three resonators of the unit cell.  
The lattice sites in the unit cell form an equilateral triangle of side $a/2$. In this paper we consider distortions of the standard Kagome lattice with two parameters: $f$ that controls the size of the triangle respect to the lattice parameter which will remain unchanged, and $\alpha$ the rotation angle of the equilateral triangle respect to its center. See Fig. \ref{kagome_lattice}.

The positions of the three sites in the unit cell are,
\begin{equation}
\vec{R}_n=f\cdot b\left ( \cos\left ( \Theta_n +\alpha\right ),\sin\left (  \Theta_n +\alpha\right ) \right )
\end{equation}
where $b=\frac{a}{2\sqrt{3}}$, $\Theta_n=n\frac{\pi}{3}-\frac{7\pi}{6}$ and $n$ labels the lattice sites $n=\{1,2,3\}$. The undistorted Kagome lattice is defined for $f=1$ and $\alpha=0$. 

\begin{figure}
\epsfig{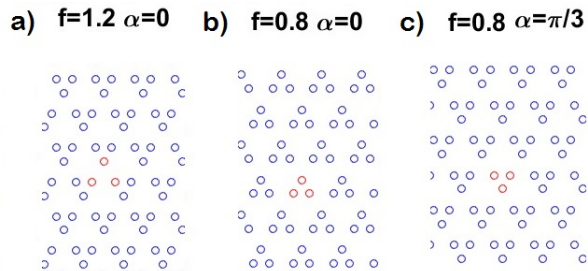}
\caption{(Color online) Real space arrangement of resonators for several deformation parameters. The unit cell is highlighted. Notice a) and c) look similar but the chosen cell is different. Notice the breaking of spatial inversion symmetry is the three cases.
 }
\label{real_space_f}
\end{figure}

Kagome lattice in our parameter space have several symmetries. For a constant $f$, there are three equivalent lattices for every $\alpha$ corresponding to $\{\alpha,\alpha+\frac{2\pi}{3},\alpha-\frac{2\pi}{3}\}$, meaning all systems in this parameter space have $C_3$ symmetry. 
For a given angle and $f<1$, the lattice with $f'=2-f$ is equivalent as well. 
However, lattices with $f<1$ and $2>f>1$ are distinguished by triangles pointing in opposite directions as shown in Fig. \ref{real_space_f} (a) and (b). Playing with parameters it is possible to create subtle differences in lattice structure, as shown in Fig. \ref{real_space_f} (a) and (c). The arrangement of resonators is the same but the unit cell where each resonator belongs are different in each case. Such configurations are therefore physically indistinguishable.
The undistorted Kagome lattice $f=1$ and $\alpha=0$ have $C_6$ symmetry, inversion symmetry  both with centers in the middle of hexagons, $C_3$ symmetry with center in the middle of triangles and three mirror symmetries. The elastic systems have time reversal symmetry as well.
 The interrelation of all these symmetries give many interesting features and we will explore some of them.

\begin{figure}
\epsfig{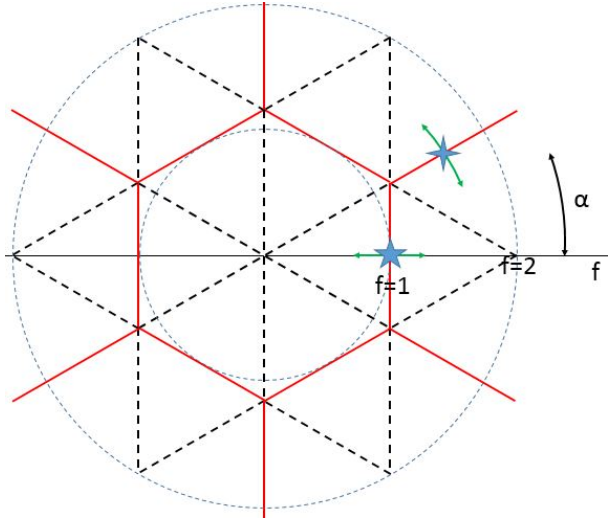}

\caption{(Color online) Gap closings in parameter space at $K$ points. Red full lines are the closing of the first gap. Dashed black lines are the closings of the second gap. The second gap is a partial gap in k-space. Topological transitions studied in this paper are marked with a five and four-pointed stars. The driving parameters are $f$ and $\alpha$ respectively and the symmetry breaking is spatial inversion and mirror symmetry respectively.
 }
\label{gaps}
\end{figure}

Due to the symmetries of the lattice, some qualitative band features are independent of the system (springs or plates). For instance, the gap closings at $K$ point of the Brillouin Zone will be relevant through the article and they are represented in Fig. \ref{gaps} in parameter space. Each red and dashed line correspond to a gap closing in cone-like shape. At momentum $K$ there are Dirac points, and opening the gap gives rise to interesting phenomena.

Spring-mass model approach is being used in Kagome lattice to explain band inversion topology and they constitute a first step towards topology in plates.

\section{Inversion symmetry breaking and topology}
\label{Sec:IV}
\subsection{Spring-mass model}
We design a spring-mass model where masses are located at sites of the Kagome lattice, i.e circles in Fig. \ref{kagome_lattice} and each blue line connecting neighboring masses are springs. The masses have only one degree of freedom, they move in the direction perpendicular to the plane. The three springs inside the unit cell have spring constant $\kappa_1$ and the springs connecting neighboring unit cells have constant $\kappa_2$. 
The equations of motion read,
\begin{equation}
\begin{matrix}
m\ddot{u}_1&=&-\kappa_1\left ( u_1-u_2 \right )-\kappa_1\left ( u_1-u_3 \right )-\kappa_2(u_1-u_2e^{-i\vec{k}\cdot\vec{a_1}})-\kappa_2(u_1-u_3e^{-i\vec{k}\vec{a_2}})\\ 
m\ddot{u}_2&=&-\kappa_1\left ( u_2-u_1 \right )-\kappa_1\left ( u_2-u_3 \right )-\kappa_2(u_2-u_1e^{i\vec{k}\vec{a_1}})-\kappa_2(u_2-u_3e^{-i\vec{k}\vec{a_3}})\\ 
m\ddot{u}_3&=&-\kappa_1\left ( u_3-u_2 \right )-\kappa_1\left ( u_3-u_1 \right )-\kappa_2(u_3-u_2e^{i\vec{k}\vec{a_3}})-\kappa_2(u_3-u_1e^{i\vec{k}\vec{a_2}})
\end{matrix}
\end{equation}
where $\vec{a}_3=\vec{a}_2-\vec{a}_1$.
Solving the temporal part as a harmonic function $u_1(t)=u_1 e^{i\omega t}$ and introducing the dimensionless $\beta=\frac{\kappa_1-\kappa_2}{\kappa_1+\kappa_2}$ which plays a role analogous to the distortion $f$ of the preceding section, and $\Omega^2=2m\omega^2/(\kappa_1+\kappa_2)$ the equation of motion reads:
\begin{equation}
-\Omega^2\begin{pmatrix}
u_1\\ u_2\\ u_3
\end{pmatrix}= \left[
\begin{pmatrix}
-4 & 1+e^{-i\vec{k}\vec{a_1}} &1+e^{-i\vec{k}\vec{a_2}} \\ 
1+e^{i\vec{k}\vec{a_1}} & -4 & 1+e^{-i\vec{k}\vec{a_3}}\\ 
1+e^{i\vec{k}\vec{a_2}} & 1+e^{i\vec{k}\vec{a_3}} & -4
\end{pmatrix}+\beta\begin{pmatrix}
0 & 1-e^{-i\vec{k}\vec{a_1}} &1-e^{-i\vec{k}\vec{a_2}} \\ 
1-e^{i\vec{k}\vec{a_1}} & 0 & 1-e^{-i\vec{k}\vec{a_3}}\\ 
1-e^{i\vec{k}\vec{a_2}} & 1-e^{i\vec{k}\vec{a_3}} & 0
\end{pmatrix}\right]
\begin{pmatrix}
u_1\\ u_2\\ u_3
\end{pmatrix}
\label{H1flexural}
\end{equation}

For $\beta=0$, we recover the dispersion relation of the undistorted Kagome lattice (analogous to $f=0$) with Dirac cones at $K$ and $K'$ points of the Brillouin Zone. Two bands cross linearly at Dirac frequency and the third band has larger energy. For $\beta\neq 0$ the gap opens up at $K$ and $K'$ points, gapping the system. Because $C_3$ is a symmetry of the lattice, its eigenvectors are eigenvectors of the system. $C_3$ rotation center located in the middle of the triangle of the unit cell gives the following matrix form for $C_3$ symmetry,
\begin{equation}
\hat{C}_3=\begin{pmatrix}
0 & 1 & 0\\ 
0 & 0 & 1\\ 
1 & 0 & 0
\end{pmatrix}
\end{equation}
Thus, the eigenvalues are $\{1,e^{i\frac{2\pi}{3}},e^{-i\frac{2\pi}{3}}\}$ and its corresponding eigenvectors, 
\begin{equation}
\begin{matrix}
u_{C_3}^{0}=\frac{1}{\sqrt{3}}\begin{pmatrix} 1\\1\\1\end{pmatrix}, & 
u_{C_3}^{+}=\frac{1}{\sqrt{3}}\begin{pmatrix} e^{-i\frac{\pi}{3}}\\e^{i\frac{\pi}{3}}\\-1\end{pmatrix},& 
u_{C_3}^{-}=\frac{1}{\sqrt{3}}\begin{pmatrix} e^{i\frac{\pi}{3}}\\e^{-i\frac{\pi}{3}}\\-1\end{pmatrix}
\end{matrix}
\label{eig_C3}
\end{equation}

These eigenvectors diagonalize the dynamical matrix (which plays the same role than a Hamiltonian) for $\beta\neq0$ at $K$ and $K'$ points. For $\beta=0$, the two states degenerate at $K$ ($K'$) point are $u_{C_3}^{0}$ and $u_{C_3}^{+}$ ($u_{C_3}^{-}$). For $\beta>0$, i.e $\kappa_1>\kappa_2$, $u_K^+=u_{C_3}^{+}$ and $u_K^-=u_{C_3}^{0}$ and reversed for $\beta<0$: $u_K^+=u_{C_3}^{0}$ and $u_K^-=u_{C_3}^{+}$. Due to the three mirror symmetries, each $K$ point is related to $K'=-K$, and its eigenvectors are the mirror symmetric of $u_K$. 
Notice that the superindex indicates different things depending on the subindex. When the subindex makes reference to $K$ point, the plus and minus signs correspond to the bands above and below the Dirac energy. The subindex $C_3$ refers to the symmetry and the plus minus or zero superindex correspond to its eigenvalues.

\begin{figure}
\epsfig{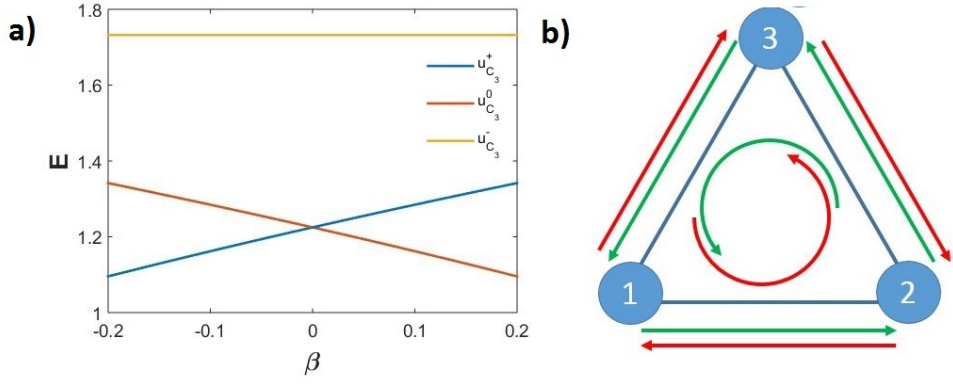}
\caption{(Color online) a) Energy level crossing at $K$ point for model in Eq. \ref{H1flexural} as a function of $\beta$. b) Representation of $M$ operator in Eq. \ref{EqM}. The arrows indicates how each component transforms, the color indicates different signs.
 }
\label{Cartoon_helicity}
\end{figure}

We see there is a crossing of eigenvectors at $\beta=0$ (the gap must close at the transition), see Fig. \ref{Cartoon_helicity} a). To capture the basic topology of this system, let's derive an effective model near each valley. The effective model can be written in the basis of the crossing eigenvectors as follows,
\begin{equation}
D_{K,ij}=\braket{u_K^i|H|u_K^j}
\label{H_eff_general}
\end{equation}
where $i,j=\{+,-\}$, we expand the dynamical matrix near each valley K and K'   the result is,
\begin{equation}
D_\eta=\begin{pmatrix}
1.5(1-\beta) & v_D(-\eta k_x+ik_y)\\ 
v_D(-\eta k_x-ik_y) & 1.5 (1+\beta)
\end{pmatrix}
\label{H_eff_T1}
\end{equation}
where $v_D=\frac{\sqrt{3}}{4}a$ and $\eta=\pm1$ for $\pm K$ and  $\vec{k}$ is measured from each valley  $\vec{k}=(\pm\frac{4\pi}{3a}+k_x,k_y)$. This is a well known model for graphene with a staggered potential \citep{effTRSB_Guinea,spinandvalleyChernnumbers,topo2Dreview}. The whole system has time reversal symmetry, one valley is transformed into the other with a time reversal transformation. However, each valley is independent from one another, since there are not direct scattering terms coupling them. Separately each valley dynamical matrix effectively behaves as a Chern insulator with broken time reversal symmetry \cite{effTRSB_Guinea} where $\beta$ is a symmetry breaking term responsible for the topological gap, and analogous to the magnetic fiel in quantum Hall phases. Opposite Chern numbers are computed near each valley when $\beta\neq 0$. Since the valleys are disconnected, a well defined valley Chern number arises. The total system is still time reversal symmetric and therefore total Chern number is zero. 

Notice that the eigenvalues of the dynamical matrix in Eq. \ref{H_eff_general}-\ref{H_eff_T1} are the square of the actual normalized frequencies $\Omega$ as in the Hamiltoninan in Eq. \ref{H1flexural}. In any case, the eigenvectors (or normal modes) and conclusions about topology hold.

We can compute the subspace generated by the two Dirac crossing vectors,
\begin{equation}
M=\ket{u_{C_3}^0}\bra{u_{C_3}^0}-\ket{u_{C_3}^+}\bra{u_{C_3}^+}=\frac{1}{2}\left(
\begin{pmatrix}
0 & 1 & 1\\ 
1 & 0 & 1\\ 
1 & 1 & 0
\end{pmatrix}-\frac{i}{\sqrt{3}}
\begin{pmatrix}
0 & 1 & -1\\ 
-1 & 0 & 1\\ 
1 & -1 & 0
\end{pmatrix}\right)
\end{equation}
This matrix corresponds to the gap-opening operator in the low energy model and is proportional to the linear term in the perturbation evaluated at $K$ point and its imaginary part is schematically represented in Fig.\ref{Cartoon_helicity} b). It gives the spatial inversion symmetry breaking term in the full spring-mass model.


This result is relevant for plates with attached resonators. The strength of springs is modeled by the distance between resonators. In our case, $\beta>0$ means that $\kappa_1$ is stronger and in a plate system is analogous to a contraction of the sites' distance in the unit cell, i.e, $f<1$. In the same way, $\beta<0$ is analogous to $f>1$.

\subsection{Plate model and valley Chern number}
\begin{figure}
\epsfig{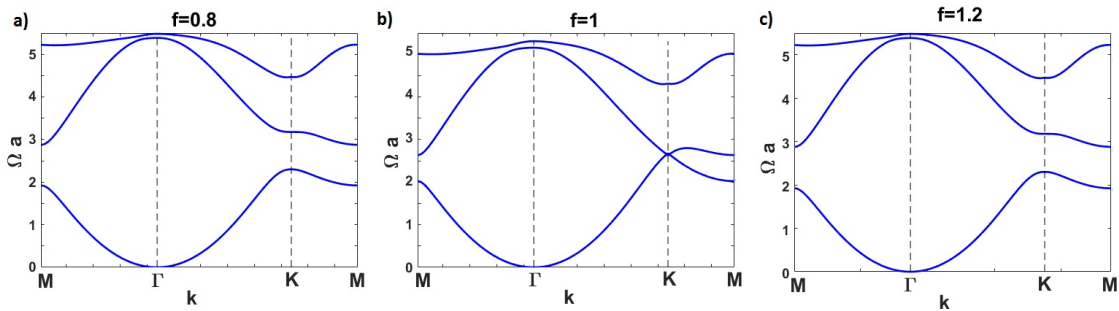},
\caption{(Color online) Band structure in a path of the hexagonal Brillouin Zone for several $f$-values and $\alpha=0$.
 }
\label{band_structure_alpha0}
\end{figure}

To reproduce previous results from spring-mass systems, we study plates with a Kagome arrangement of resonators and model spring strength with distance between resonators. Fixing $\alpha=0$ and varying $f$ around $1$ let us model the variation of spring constants within the unit cell ($\kappa_1$) respect to the springs connecting different cells ($\kappa_2$). The corresponding band structures are in Fig. \ref{band_structure_alpha0}. At both sides of the transition the band structure is the same (see their similar spatial distribution in Fig. \ref{real_space_f} a) and c)). However, topology encoded in eigenvectors is inverted as we will see. At the transition point $f=1$, the two bands form Dirac cones at first order in momentum around $K$ and $K'$. The Dirac energy is $\Omega_D a=2.5$.

\begin{figure}
\epsfig{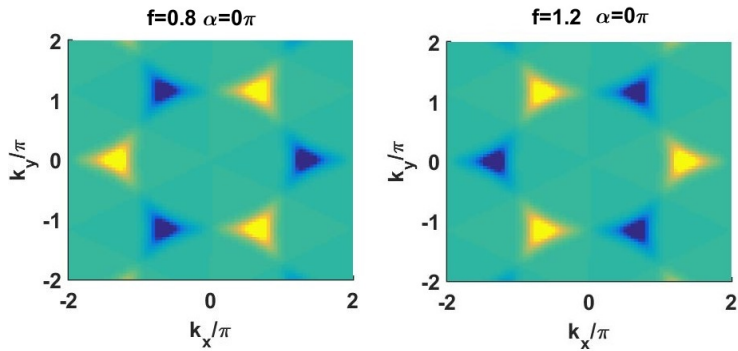}
\caption{(Color online) Berry curvature of the lower band over the first Brillouin zone. Berry curvature is localized at $K$ and $K'$ points with different signs for different phases. Blue is negative and yellow is positive.
 }
\label{BerryCurvature_T1}
\end{figure}

For $f\neq 1$ spatial inversion symmetry is broken, while the remaining symmetries are still present (See Fig. \ref{real_space_f}). The broken inversion symmetry allows us to define a valley Chern number, as previously stated in the spring-mass model. 
In Fig. \ref{BerryCurvature_T1} the computed Berry curvature of first band is plotted. 
The Berry Curvature in 2D $k$-space is,
\begin{equation}
B=-i
\braket{\partial_x u_k|\partial_y u_k}+i
\braket{\partial_y u_k|\partial_x u_k}
\label{BerryCurvature}
\end{equation}
where $u_k$ is the eigenvector of one band at momentum $k$. The eigenvector is computed from the PWE method as the null space of $A$ matrix in Eq. \ref{Eq_bandsDani}. We observe that the Berry curvature is localized near $K$ and $K'$ with opposite sign and it changes at the transition. 

\begin{figure}
\epsfig{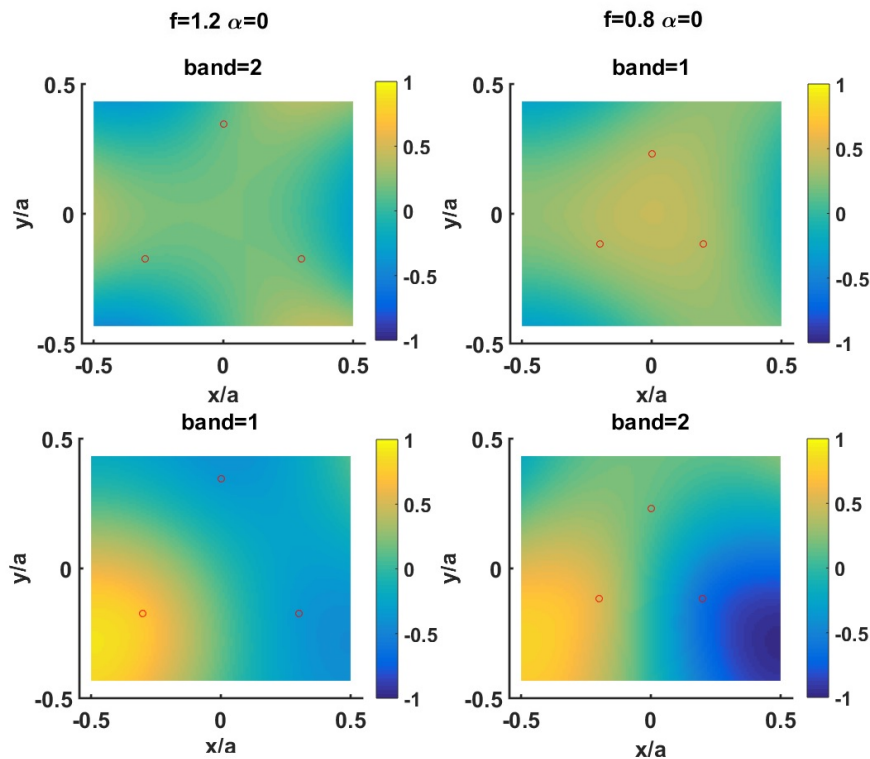}
\caption{(Color online) Mode shapes. Real part of $w(\vec{r})$ for different bands and phases. Notice the analogy of the first row with the eigenvalue of the spring model $u_{C_3}^0=\frac{1}{\sqrt{3}} (1,1,1)^t$ or in the second row with the real part $\textup{Re}\{u_{C_3}^+\}=\frac{1}{\sqrt{3}} (0.5,0.5,-1)^t$. Notice the band inversion. Mode shapes are not periodic due to the phase $e^{-i\vec{K}\cdot\vec{r}}$ in Eq. \ref{Eq_eigenstates}.
 }
\label{Eig_1}
\end{figure}

For further analogy with the spring system, we compute the mode shapes in real space at the $K$ point for the two lower bands in Fig. \ref{Eig_1} which closely resemble the eigenvectors involved in the transition $u_K^\pm$. Moreover, band inversion is clearly seen. The mode shapes switch energies at both sides of the transition in the same way than eigenvectors in the spring-mass model (Fig. \ref{Cartoon_helicity} a).


\subsection{Edge states in ribbons}
\begin{figure}
\epsfig{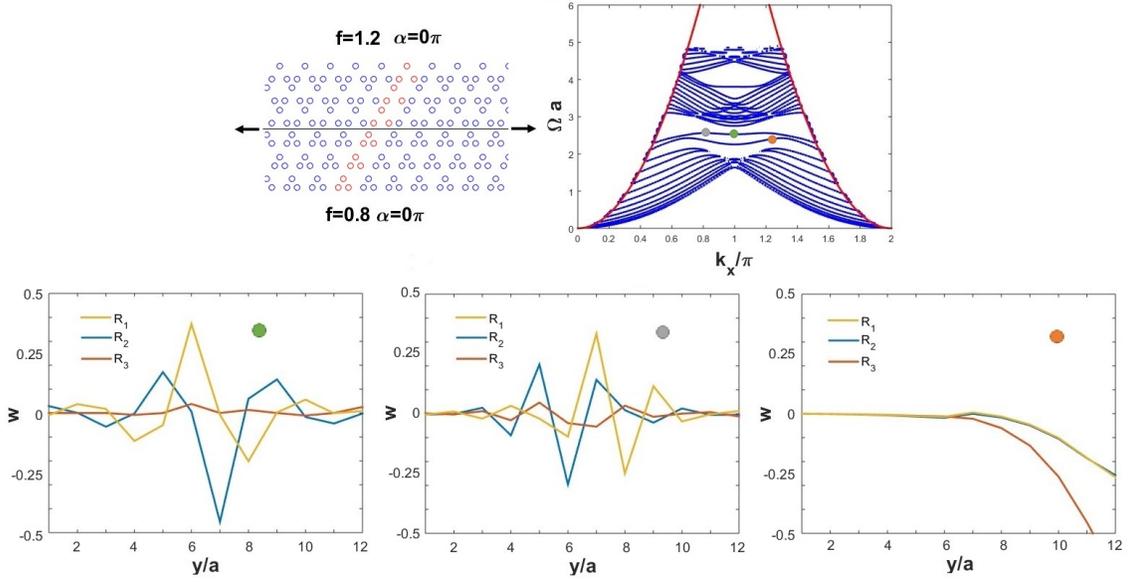}\\

\caption{(Color online) Ribbon of resonators over an infinite plate. The system contains a domain wall between two phases with opposite valley Chern numbers. At the top left, real space ribbon representation. The horizontal line separates the two phases and black arrows indicate that the ribbon is infinite in horizontal direction. In red, resonators in one supercell. At the top right there is the band structure of the finite system, neglecting non-bulk modes, i. e. modes in the interior of the free dispersion curve $\Omega a=\left( \frac{k_x}{\pi}\right)^2$. Two mid gap bands appear. At the bottom, mode shapes or in other words, real space displacement field along the supercell sites $w(\vec{R}_\alpha)$ for different frequencies and momenta as indicated with colored dots on the band structure.
 }
\label{DomainWall1}
\end{figure}

\begin{figure}
\epsfig{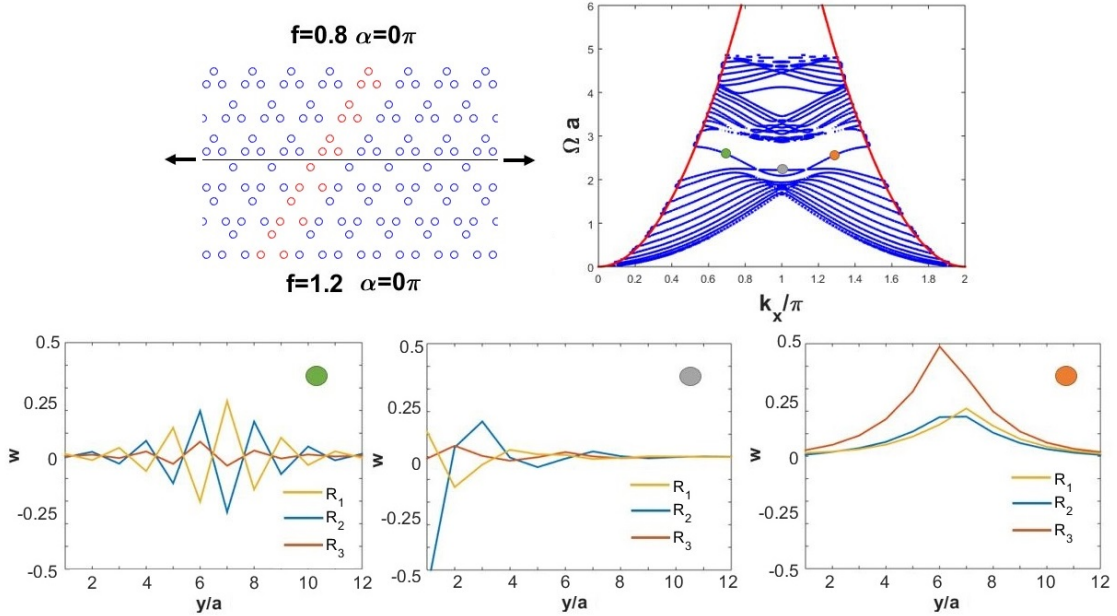}\\
\caption{(Color online) Ribbon of resonators over an infinite plate. The system contains a domain wall between two phases with opposite valley Chern numbers (See Fig. \ref{BerryCurvature_T1}). At the top left, real space ribbon representation. The horizontal line separates the two phases and black arrows indicate that the ribbon is infinite in horizontal direction. In red, resonators in one supercell. At the top right there is the band structure of the finite system, neglecting non-bulk modes, i. e. modes in the interior of the free dispersion curve $\Omega a=\left( \frac{k_x}{\pi}\right)^2$. One mid gap band appears. At the bottom, mode shapes or in other words, real space displacement field along the supercell sites $w(\vec{R}_\alpha)$ for different frequencies and momenta as indicated with colored dots on the band structure.
 }
\label{DomainWall2}
\end{figure}

In this section we study the interface states appearing between two lattices with distinct valley Chern numbers, which are topologically protected \citep{spinandvalleyChernnumbers,topo2Dreview,Ruzzene_graphene} i. e. with zig-zag interfaces. For analogy with graphene-like lattices, we call zig-zag edges those that go along directions $\vec{a}_1$, $\vec{a}_2$ or $\vec{a}_2-\vec{a}_1$. We call armchair interface in Kagome lattice to the one along vertical direction $\vec{a}_2-\frac{1}{2}\vec{a}_1$ in our definition of the unit cell.
We create ribbons in a supercell along $\vec{a}_2$ direction and periodic in $\vec{a}_1$ direction. Even ribbons with valley topological phases in electronic system don't have gapless edges states, because valleys are not well defined in vacuum, unless the boundary is with another topological phase with opposite valley Chern number \cite{spinandvalleyChernnumbers}. The same reasoning is true for plates. Therefore, boundary states appear at the interface between two phases with opposite signed topological invariants. Such interface is contained in the supercell of the ribbons as shown in Fig \ref{scketch_ribbon}. Two types of interfaces can be made, which are depicted in Fig. \ref{DomainWall1} and \ref{DomainWall2}. Schematic real space supercell is highlighted, a black full line separates two topological phases distinguished by opposite valley Chern numbers. 
The bands are limited by the free-wave dispersion relation, outside that region there are not bulk solutions of the plate equation. The two types of interfaces exhibit a band of boundary states localized at the domain wall. In Fig. \ref{DomainWall1} a second band appears containing edge states at the top of the ribbon which are non-topological. An analogous band is present in Fig. \ref{DomainWall2} with edge states at the bottom of the ribbon as can be seen in the mode shapes.

\begin{figure}
\epsfig{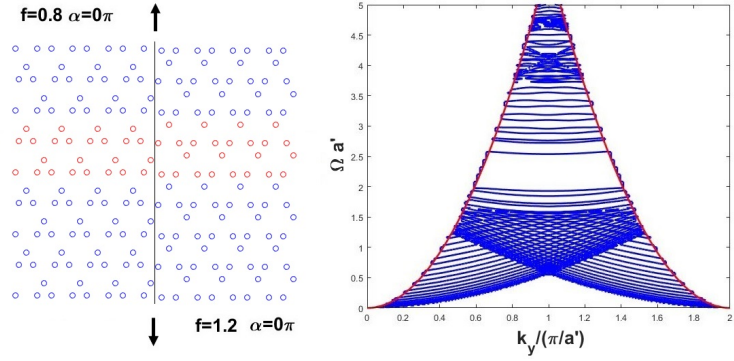}\\
\caption{(Color online) Ribbon of resonators over an infinite plate. The system contains a domain wall between two phases with opposite valley Chern numbers (See Fig. \ref{BerryCurvature_T1}). On the left, real space ribbon representation. The vertical line separates the two phases and black arrows indicate that the ribbon is infinite in vertical direction. The interface is armchair-like. The band structure does not show localized modes within gap frequencies, bands that appear isolated at gap frequencies are bulk modes.
 }
\label{DomainWall1_vertical}
\end{figure}

The topological edge modes are robust against certain types of perturbations that do not mix valleys. We have confirmed this fact by corroborating that these states are not removed away by the addition of general perturbations to the boundary. However, there are perturbations mixing valley degrees of freedom such an armchair boundary \cite{edges_photonic} that will destroy the protection as can be seen in Fig. \ref{DomainWall1_vertical}. Notice the change in the unit cell parameter, now in the direction of periodicity it is $a'=\sqrt{3}a$.


\subsection{Finite systems}

\begin{figure}
\epsfig{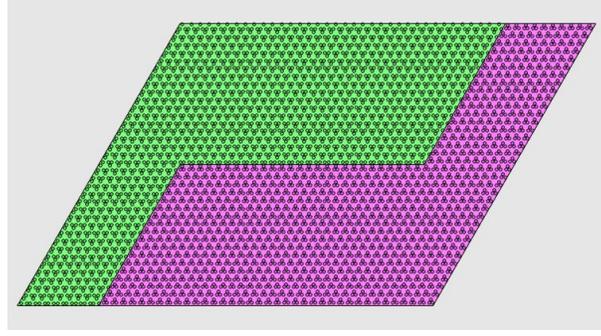}\\
\caption{(Color online) Schematic representation of a cluster of resonators on top of an infinite plate. The cluster is designed with a Z-shaped interface.
}
\label{scketch_cluster}
\end{figure}

\begin{figure}
\epsfig{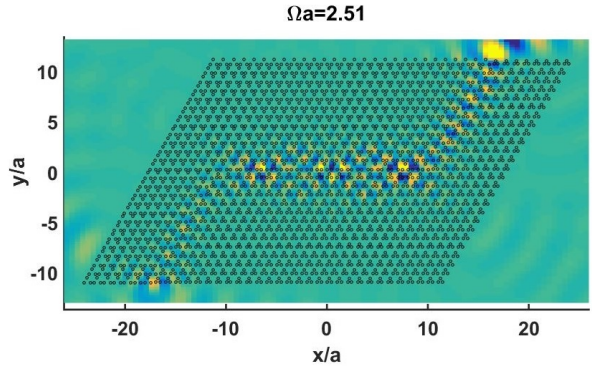}
\caption{(Color online) MST simulations of an arrangement of resonators with two phases separated by a domain wall in zig-zag. The frequency is tuned so the mode is in a gap and correspond to topological edge states.
 }
\label{Fig_MST}
\end{figure}

Now, we study a finite cluster of resonators on top of an infinite plane where multiple scattering theory described in section II and developed in Ref. \citep{Dani_graphene} applies. The cluster of resonators contain two phases separated by a zig-zag interface with Z-shape, Fig. \ref{scketch_cluster}. Topological protected state appears at mid gap energy. Notice that the horizontal interface is equivalent to the domain wall in Fig. \ref{DomainWall1}, thus the frequency is tuned to find topological edge modes, in this case $\Omega a=2.51$. 
Fig. \ref{Fig_MST} shows an edge state without backscattering, this mode is being computed without external input field, i.e. $\psi_0=0$. The vector of coefficients $\psi_e(\vec{R}_\beta)$ in Eq.\ref{Eq_MST2} is the right-singular vector whose single value is zero. This method computes natural excitations of the system at a given frequency. 

\begin{figure}
\epsfig{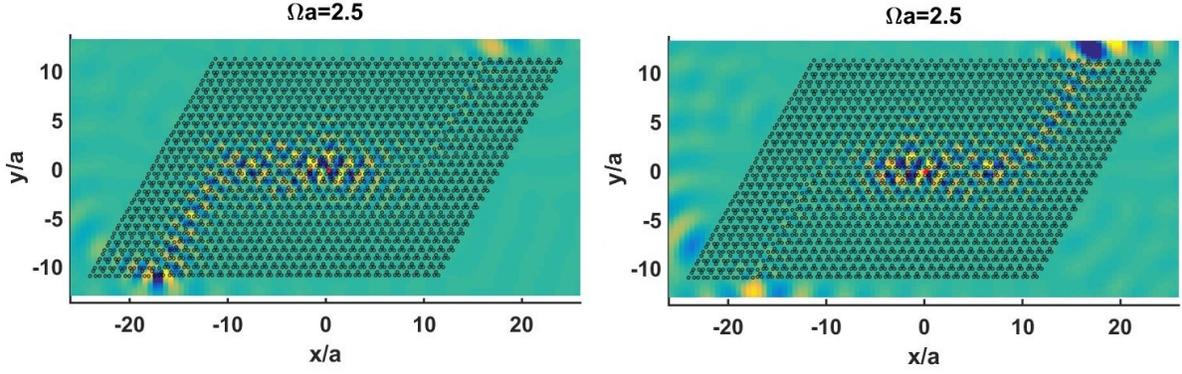}
\caption{(Color online) MST simulations of an arrangement of resonators with two phases separated by a zig-zag domain wall (no mixing valleys) in Z-shape. The frequency is tuned so the modes are  topological edge states. The red dots correspond to the two excitation points $\vec{x}_1=(-1,0)a$ and $\vec{x}_2=(1,0)a$. The temporal dephasing is $\varphi=0$ (the two points are excited simultaneously) and $\varphi=\pi$ (anti-phase excitation) respectively.
 }
\label{Fig_MST_T1}
\end{figure}

Moreover, in the same cluster we find appropriate multipoint excitation with dephasing in time. A two-point excitation $\psi_0(\vec{R}_\alpha)=G_0(\vec{R}_\alpha-\vec{x}_1)+G_0(\vec{R}_\alpha-\vec{x}_1)e^{i\varphi}$ where point sources are located at the horizontal domain wall, $\vec{x}_1=(-1,0)a$ and $\vec{x}_2=(1,0)a$. The dephasing $\varphi$ is varied until propagating waves in one direction only are tuned. The results are shown in Fig. \ref{Fig_MST_T1} and are similar to those presented in Ref. \cite{ccy}.


\section{Mirror symmetry breaking and topology}
\label{Sec:V}
\subsection{Spring-mass model}
\begin{figure}
\epsfig{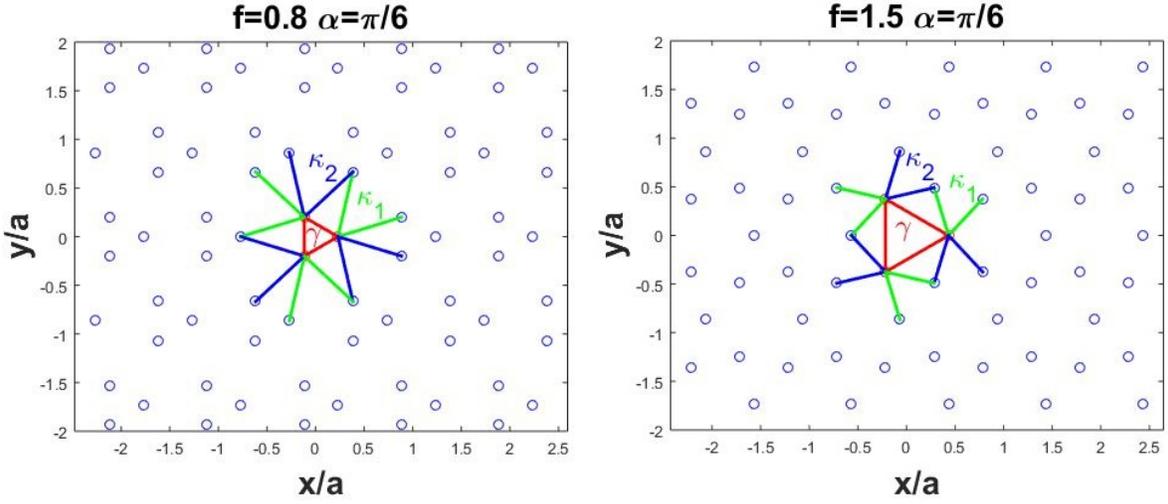}
\caption{(Color online) Distorted Kagome lattice for two $f$ values and $\alpha=\pi/6$. Notice that spatial inversion is not a symmetry of the system. Increasing slightly $\alpha$ shortens green links and enlarges blue links. The spring model models changes in distance with appropriate changes in $\beta$.
 }
\label{spring_pi6}
\end{figure}

\begin{figure}
\epsfig{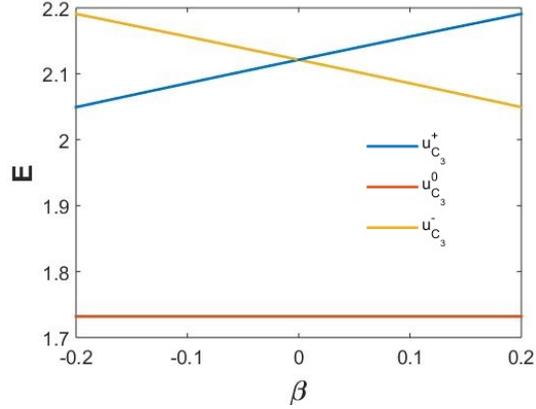}
\caption{(Color online) Energy levels at $K$ point for model in Eq. \ref{H_T2} as a function of $\beta$ for $\gamma'=1$.
 }
\label{beta_T2}
\end{figure}

Now we consider a model with mirror symmetry at $\alpha=\frac{\pi}{6}$ and consider two continuous deformations that break mirror symmetry. Changing $\alpha$ towards one side or the other will give two phases differentiated by different eigenvectors of $C_3$ symmetry. The spring-mass model is constructed by changing the relative spring constant between green and blue springs as indicated in Fig. \ref{spring_pi6}.
The equation of motion read,
\begin{equation}
\begin{matrix}
m\ddot{u}_1&=&-\gamma\left ( u_1-u_2 \right )-\gamma\left ( u_1-u_3 \right )-\kappa_1(u_1-u_2e^{-i\vec{k}\vec{a_2}})-\kappa_1(u_1-u_3e^{-i\vec{k}\vec{a_3}})-\kappa_2(u_1-u_2e^{-i\vec{k}\vec{a_1}})-\kappa_2(u_1-u_3e^{-i\vec{k}\vec{a_2}})\\ 
m\ddot{u}_2&=&-\gamma\left ( u_2-u_1 \right )-\gamma\left ( u_2-u_3 \right )-\kappa_1(u_2-u_1e^{i\vec{k}\vec{a_2}})-\kappa_1(u_2-u_3e^{i\vec{k}\vec{a_1}})-\kappa_2(u_2-u_1e^{i\vec{k}\vec{a_1}})-\kappa_2(u_2-u_3e^{-i\vec{k}\vec{a_3}})\\ 
m\ddot{u}_3&=&-\gamma\left ( u_3-u_2 \right )-\gamma\left ( u_3-u_1 \right )-\kappa_1(u_3-u_2e^{-i\vec{k}\vec{a_1}})-\kappa_1(u_3-u_1e^{i\vec{k}\vec{a_3}})-\kappa_2(u_3-u_2e^{i\vec{k}\vec{a_3}})-\kappa_2(u_3-u_1e^{i\vec{k}\vec{a_2}})
\end{matrix}
\end{equation}
introducing the relative difference $\beta=\frac{\kappa_1-\kappa_2}{\kappa_1+\kappa_2}$ we rewrite the system of equations in matrix form,
\begin{equation}
\begin{matrix}
-\Omega^2
\begin{pmatrix}
u_1\\ u_2\\ u_3
\end{pmatrix}= 
\gamma'
\begin{pmatrix}
-2 & 1 & 1\\ 
1 & -2 & 1\\
1 & 1 & -2
\end{pmatrix}
\begin{pmatrix}
u_1\\ u_2\\ u_3
\end{pmatrix}+
\begin{pmatrix}
-4 & e^{-i\vec{k}\vec{a_2}}+e^{-i\vec{k}\vec{a_1}} & e^{-i\vec{k}\vec{a_3}}+e^{-i\vec{k}\vec{a_2}}\\ 
e^{i\vec{k}\vec{a_2}}+e^{i\vec{k}\vec{a_1}} & -4 & e^{i\vec{k}\vec{a_1}}+e^{-i\vec{k}\vec{a_3}}\\ 
e^{i\vec{k}\vec{a_3}}+e^{i\vec{k}\vec{a_2}} & e^{-i\vec{k}\vec{a_1}}+e^{i\vec{k}\vec{a_3}} & -4
\end{pmatrix}\begin{pmatrix}
u_1\\ u_2\\ u_3
\end{pmatrix}
\\
+\beta\begin{pmatrix}
0& e^{-i\vec{k}\vec{a_2}}-e^{-i\vec{k}\vec{a_1}} & e^{-i\vec{k}\vec{a_3}}-e^{-i\vec{k}\vec{a_2}}\\ 
e^{i\vec{k}\vec{a_2}}-e^{i\vec{k}\vec{a_1}} & 0 & e^{i\vec{k}\vec{a_1}}-e^{-i\vec{k}\vec{a_3}}\\ 
e^{i\vec{k}\vec{a_3}}-e^{i\vec{k}\vec{a_2}} & e^{-i\vec{k}\vec{a_1}}-e^{i\vec{k}\vec{a_3}} & 0
\end{pmatrix}\begin{pmatrix}
u_1\\ u_2\\ u_3
\end{pmatrix}
\end{matrix}
\label{H_T2}
\end{equation}
where $\Omega^2=2m\omega^2/(\kappa_1+\kappa_2)$ and $\gamma'=2\gamma/(\kappa_1+\kappa_2)$
The eigenvectors at $K$ point are the same eigenvectors of $C_3$ symmetry, but its energy order is different from previous section. Now, $u_K^\pm=u_{C_3}^{\pm}$, i. e. gap closes at $K$ point forming a the Dirac cone at $\alpha=\frac{\pi}{6}$. Notice the closing occurs on first or second gap depending on $f$ (See Fig. \ref{gaps}). In any case, the Dirac cones are made of states with complex conjugate eigenvalues of $C_3$ symmetry. Moreover, they are interchanged at the transition: $u_K^\pm=u_{C_3}^{\pm}$ for $\beta>0$ and $u_K^\pm=u_{C_3}^{\mp}$ for $\beta<0$ see Fig. \ref{beta_T2}; and interchanged again at the other valley $K'$.

We compute the effective model for this band crossing system as in Eq. \ref{H_eff_general}. The result is,
\begin{equation}
D_\eta=\begin{pmatrix}
3\gamma'+1.5(1-\beta) & \eta v_D e^{i2\pi/3}(k_x+ik_y)\\ 
v_D \eta e^{-i2\pi/3}(k_x-ik_y) & 3\gamma'+1.5 (1+\beta)
\end{pmatrix}
\end{equation}
where $v_D=\frac{\sqrt{3}}{2}a$. By rotating $\vec{k}=(k_x,k_y)$-reference system by $\pi/3$, the dynamical matrix can be written with the same structure than Eq. \ref{H_eff_T1},
\begin{equation}
D_\eta=\begin{pmatrix}
3\gamma'+1.5(1-\beta) & v_D( \eta k_x'+ik_y')\\ 
v_D(\eta k_x'-ik_y') & 3\gamma'+1.5 (1+\beta)
\end{pmatrix}
\label{H_eff_T2}
\end{equation}

This result illustrates that the mirror symmetry breaking in the original model is analogous to an inversion symmetry in graphene-like systems where $\beta$ is the pseudo-magnetic field in quantum valley Hall effect. Instead of inducing nonequivalent sublattice potential, here the potential is between eigenstates of the system and $C_3$.

The subspace generated by the two Dirac eigenstates crossing at $K$ differentiates between states rotating in different directions $M=i(\hat{C}_3-\hat{C}_3^{t})$,
\begin{equation}
M=\ket{u_{C_3}^+}\bra{u_{C_3}^+}-\ket{u_{C_3}^-}\bra{u_{C_3}^-}=\frac{i}{\sqrt{3}}\begin{pmatrix}
0 & 1 & -1\\ 
-1 & 0 & 1\\ 
1 & -1 & 0
\end{pmatrix}
\label{EqM}
\end{equation} 
This matrix is proportional to the linear term in $\beta$ at $K$ point and it is schematically represented in Fig.\ref{Cartoon_helicity} b). This gives us the mirror symmetry breaking effect in real space lattice vectors.

\subsection{Plate model and valley Chern number}
\begin{figure}
\epsfig{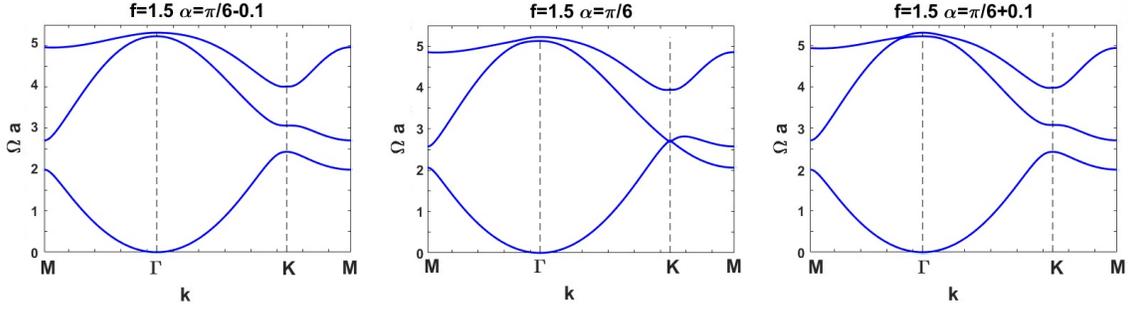}
\caption{(Color online) Band structure of deformed Kagome lattice for several $\alpha$-values and $f=1.5$.
 }
\label{band_structure_f15}
\end{figure}

In this section we plot several band structures around $\alpha=\pi/6$. 
Notice that in Fig. \ref{gaps} the gap closes for all $f$ at $\alpha=\pi/6$ at $K$ point. For $f<\frac{2}{\sqrt{3}}$, the second and third bands are degenerate at $K$ point. For $f>\frac{2}{\sqrt{3}}$ the first and second bands form the Dirac cone. The two transitions have equivalent topology. In the spring-mass model this corresponds to varying the value of $\gamma$ that tunes the energy of $u_{C_3}^0$ but does not affect the other two crossing states. However, the gap opening at $K$ when $\alpha\neq\pi/6$ is not complete for small $f$. For large $f$, $K$ point is not the minimum of the second band, although topological states come from what happens at $K$ point, the gap is complete and we show the results of for $f=1.5$. The band structures of plates with different arrangements of resonators are plotted in Fig. \ref{band_structure_f15}. At equidistant points in parameter space from the transition points the band structures are the same, however their topology is not. At the transition point, a Dirac cone at $K$ point is formed which opens upon breaking mirror symmetry. The Dirac energy is $\Omega a=2.7$.

\begin{figure}
\epsfig{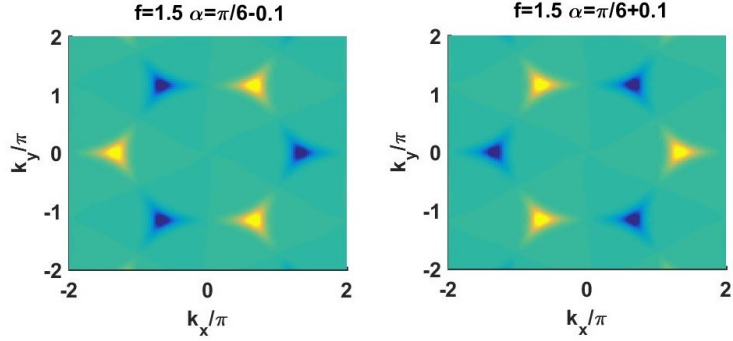}
\caption{(Color online) Berry curvature of the lower band over the first Brillouin zone. Berry curvature is localized at $K$ and $K'$ points with different signs for different phases. Blue is negative and yellow is positive.
 }
\label{BerryCurvature_T2}
\end{figure}

For $\alpha\neq \frac{\pi}{6}$ mirror symmetry is broken and as we show in the effective model we can define a Berry curvature as in Eq. \ref{BerryCurvature}.
The result is shown in Fig. \ref{BerryCurvature_T2}.
The eigenvectors in the Brillouin Zone are computed from PWE method as the null space of $A$ matrix in Eq. \ref{Eq_bandsDani} at the appropriate frequency as described in Ref. \citep{Dani_graphene}. We observe that the Berry curvature is localized near $K$ and $K'$ with opposite sign and it changes at the transition, consistently with the effective spring-mass model.

\subsection{Edge states in ribbons}
\begin{figure}
\epsfig{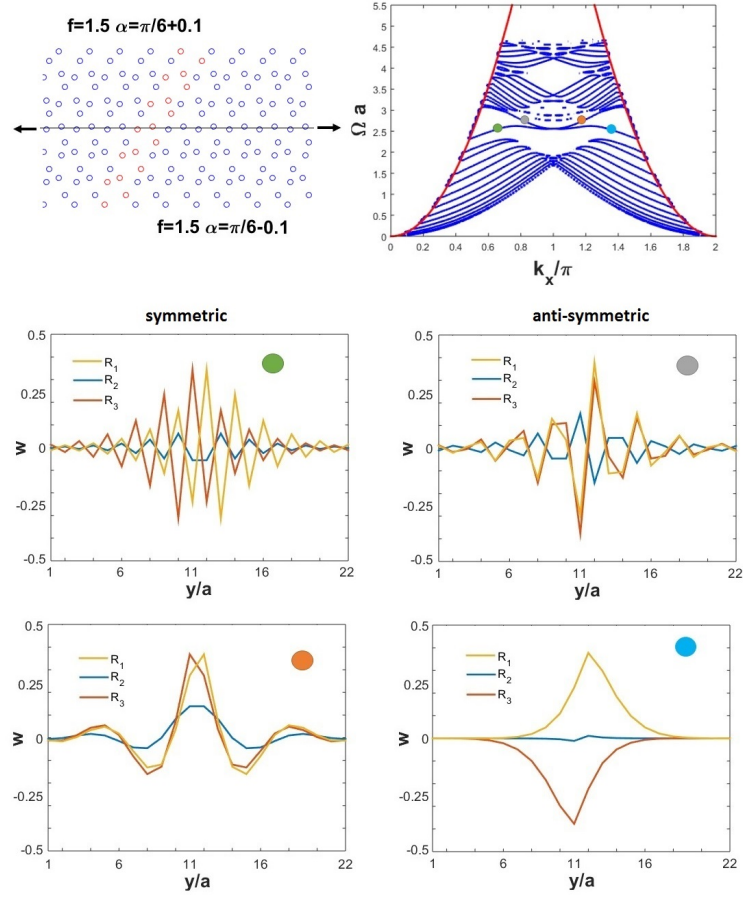}\\
\caption{(Color online)  Ribbon of resonators over an infinite plate. The system contains a domain wall between two phases with opposite valley Chern numbers (See Fig. \ref{BerryCurvature_T1}). At the top left, real space ribbon representation. The horizontal line separates the two phases and black arrows indicate that the ribbon is infinite in horizontal direction. In red, resonators in one supercell. At the top right there is the band structure of the finite system, neglecting non-bulk modes, i. e. modes in the interior of the free dispersion curve $\Omega a=\left( \frac{k_x}{\pi}\right)^2$. Two crossing bands appear in the gap, they have different symmetry under domain wall spatial inversion as seen at the bottom. At the bottom, mode shapes, i. e. real space displacement field along the supercell sites $w(\vec{R}_\alpha)$ for different frequencies and momenta as indicated with colored dots on the band structure. 
 }
\label{DomainWallT2_f15}
\end{figure}

We compute the edge states of a ribbon with an interface and find two crossing bands in the middle of the gap. The crossing indicates that the two bands have different symmetry. 
In Fig. \ref{DomainWallT2_f15}, edge states appear in the boundary of the two phases, due to the different valley Chern numbers. In this transition there are two crossing bands with different symmetries that are topologically protected. The different symmetries can be observed in the modes in Fig. \ref{DomainWallT2_f15}. They are symmetric or anti-symmetric respect to the domain wall. Notice site two maps onto itself under inversion at the domain wall and site three and one maps onto one another. This symmetry in the eigenvectors reflect the inversion symmetry present in real space in the ribbon due to the fact that phases are equidistant in real space from the transition point in parameter space. In other words, the two phases are characterized by $\alpha=\pi/6\pm \phi$, where $\alpha=\pi/6$ is the transition point and $\phi=0.1$. 
This ribbon symmetry is also present in ribbons with two phases breaking inversion symmetry in honeycomb lattice like in Ref \citep{Ruzzene_graphene}. Since the two phases are equidistant from the transition point, there is an inversion that gives symmetric and anti-symmetric edge modes respect to the domain wall. (See Appendix \ref{Appendix}). Unlike honeycomb lattice in Kagome arrangement each number site has its inversion point. In graphene, the spatial inversion is clearly seen in the eigenvectors $u_A=(1,0)^t$ and $u_B=(0,1)^t$ than transform into one another by appropriate inversion in real space. In our case, Eq. \ref{H_eff_T2}, the eigenvectors at a given frequency and at each side of the domain wall are related by spatial inversion too
\begin{equation}
\begin{matrix}
u_K^{\beta>0}=\frac{1}{\sqrt{3}}\begin{pmatrix} e^{-i\frac{2\pi}{3}}\\1\\e^{i\frac{2\pi}{3}}\end{pmatrix}= e^{i\frac{\pi}{3}}u_{C_3}^{+}& 
u_K^{\beta<0}=\frac{1}{\sqrt{3}}\begin{pmatrix} e^{i\frac{2\pi}{3}}\\1\\e^{-i\frac{2\pi}{3}}\end{pmatrix}=e^{i\frac{\pi}{3}}u_{C_3}^{-}
\end{matrix}
\end{equation}
Site 2 maps into itself, while sites 1 and 3 interchange and appropriate combinations. The result shows symmetric and anti-symmetric modes, as observed in the ribbon eigenvectors (Fig \ref{DomainWallT2_f15}). 
Notice inversion symmetry is not present in domain walls in ribbons with phases of Kagome lattice with broken inversion symmetry shown in Fig. \ref{DomainWall1}-\ref{DomainWall2}. Modes are not symmetric or anti-symmetric and neither the eigenvectors at $K$ involved in the transition ($u_{C_3}^{0}$ and $u_{C_3}^{+}$) exhibit inversion symmetry, as expected.\\

\begin{figure}
\epsfig{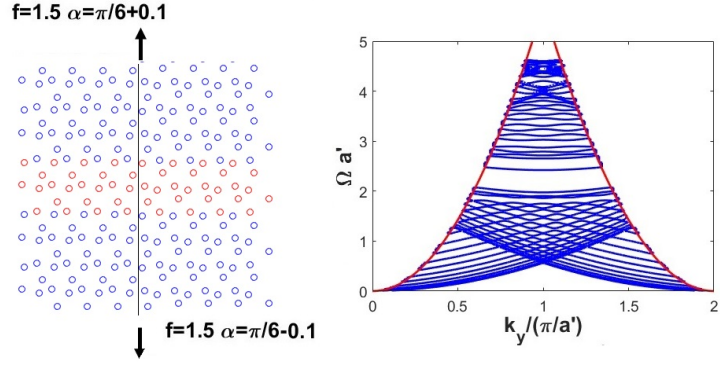}\\
\caption{(Color online) Ribbon of resonators over an infinite plate. The system contains a domain wall between two phases with opposite valley Chern numbers (See Fig. \ref{BerryCurvature_T1}). On the left, real space ribbon representation. The vertical line separates the two phases whose interface is armchair type and black arrows indicate that the ribbon is infinite in the vertical direction. The band structure does not show localized modes within gap frequencies, bands that appear isolated at gap frequencies are bulk modes.
 }
\label{DomainWallT2_vertical}
\end{figure}

Valley topology is not protected against perturbations mixing the valleys. For instance, a vertical interface, (armchair type) mixes the valleys and the edge states disappear as shown in Fig. \ref{DomainWallT2_vertical}. The bands displayed in the middle of the gap are also bulk bands.

\subsection{Finite systems}
\begin{figure}
\epsfig{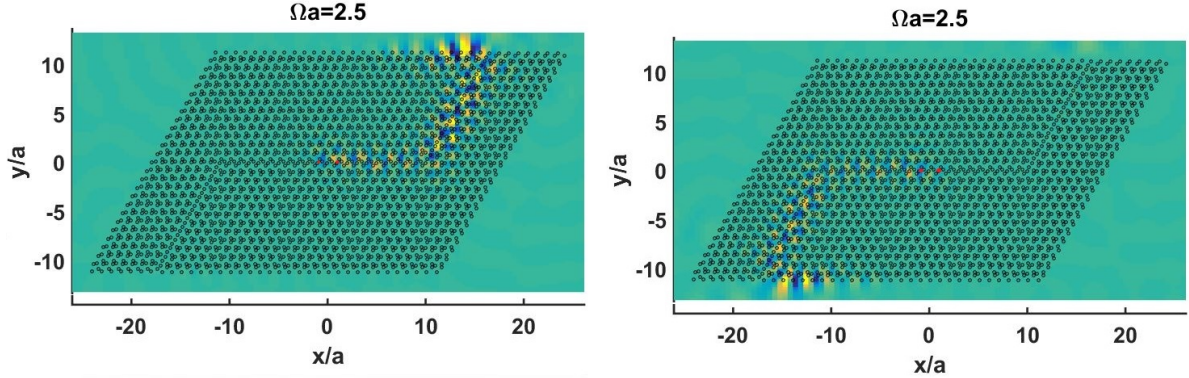}
\caption{(Color online) MST simulations of an arrangement of resonators with two phases separated by a zig-zag domain wall in Z-shape. The frequency is tuned so the mode is in a gap and correspond to topological edge states. The red dots correspond to the two excitation points. The dephasing is $\varphi=\pi$ on the left and $\varphi=-0.36\pi$ on the right.
 }
\label{Fig_MST_T2}
\end{figure}

We design a finite structure of resonators over an infinite plate and compute the real part of $w(\vec{r})$. A similar result occurs for natural modes of the system, as in Fig. \ref{Fig_MST}. We also find two-point time-dephased excitation at mid gap frequency, so one-way propagation is achieved. See Fig. \ref{Fig_MST_T2}, the red dots are the points where the external excitation force is applied, $\vec{x}_1=(-1,0)$ and $\vec{x}_2=(1,0)$. Different dephasing $\varphi$ excites different directional waves.

\section{Conclusion}
\label{Sec:VI}

We have studied two types of topological transitions in mechanical metamaterials based on the distorted Kagome lattice, namely inversion symmetry or mirror symmetry breaking. In spring-mass systems, we derived a dynamical matrix for each valley that effectively behaves as a Chern insulator. We have identified, in the microscopic model, the operator acting as a pseudo-magnetic field which is controlled by relative values of springs' strengths. We also exploit this finding for flexural waves in plates coupled to resonators. In this context the "magnetic field" is controlled by the  distance between resonators. The main manifestation of the valley Hall effect in our system is the presence of protected boundary states located at interfaces between domains with opposite signed valley Chern numbers. These interfaces must have appropriate edges as shown in simulations of ribbons and finite clusters of resonators with zig-zag domains.   
We also illustrated how mixing valleys with armachair-type interfaces produces back-scattering  and destroys the topological modes. However, we also claim that a lattice lacking inversion symmetry at the transition despite intact mirror symmetry exhibits the same type of valley topology of broken mirror symmetry. We compute a similar effective model for springs and find protected edge states with different symmetry. We find simple two-point excitation generating one-way flexural waves in finite systems that can propagate through desired bends in 2D space. It is well known that the dynamics of spring-mass systems  is dissimilar in several ways to the one of interacting resonators coupled to plates. 
For instance, interaction between the resonators is long-ranged and the dynamical matrix is frequency dependent in the latter. However, throughout this work we have established a common origin to their topological properties. We hope all these findings help enlightening the path towards future applications in wave guiding and related fields.

$\mathit{Acknowledgments}$: N.L and J.V.A. acknowledges financial support from MINECO grant FIS2015-64886-C5-5-P. NL acknowledges financial support from the
Spanish Ministry of Economy and Competitiveness, through
The "Mar\'ia de Maeztu" Programme for Units of Excellence in R\&D (MDM-2014-0377), and also hospitality from the Universitat Jaume I in Castellon where part of this work was done. D.T. acknowledges financial support through the ``Ram\'on y Cajal'' fellowship under grant number RYC-2016-21188. P.S-J. acknowledges financial support from the Spanish Ministry of Economy and Competitiveness through Grant No. FIS2015-65706-P (MINECO/FEDER) J. C. acknowledges the support from the European Research Council (ERC) through the Starting Grant No. 714577 PHONOMETA and from the MINECO through a Ram\'on y Cajal grant (Grant No. RYC-2015-17156).


\section{Appendix: honeycomb ribbons with broken inversion symmetry}
\label{Appendix}

\begin{figure}
\epsfig{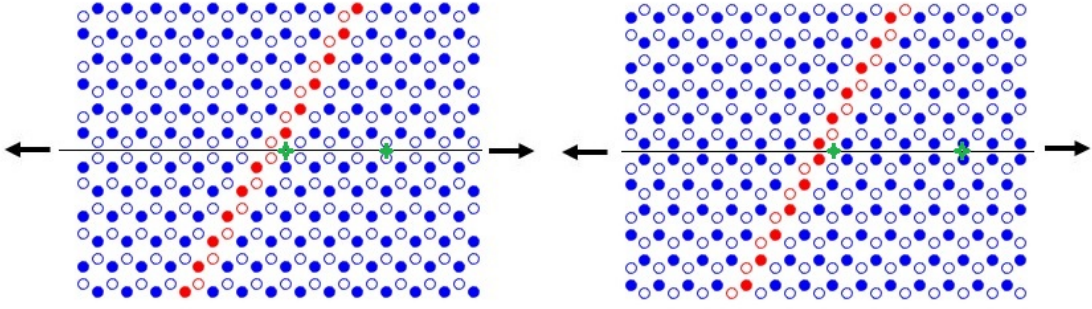}\\
\caption{(Color online) Schematic of hexagonal arrangement of resonators having two different masses (filled or empty circles represent heavy and light masses) with soft (light-light) and hard (heavy-heavy) interfaces. The black full line is the domain wall. In green star marker, the inversion symmetry centers of the ribbon. The arrows represent the infinity of the ribbon in horizontal direction.
 }
\label{graphenelattice}
\end{figure}

\begin{figure}
\epsfig{file=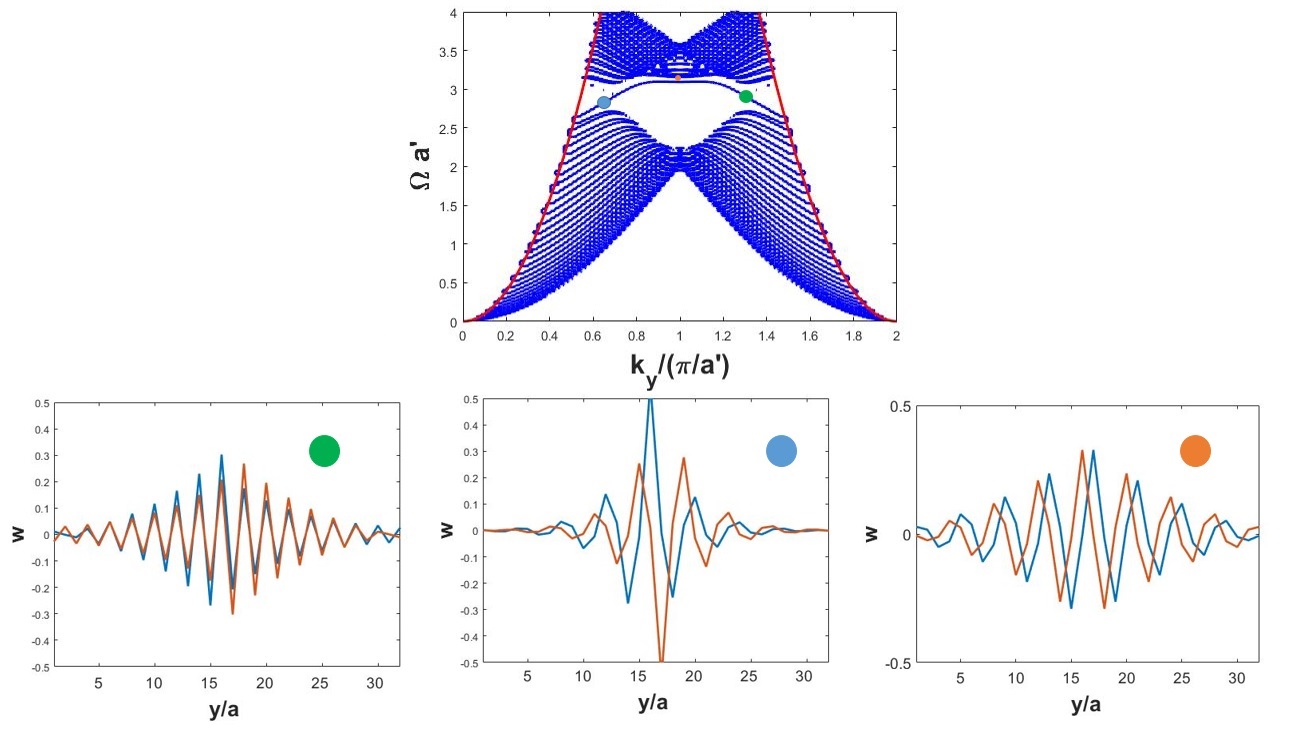,width=15cm}\\
\caption{(Color online) Ribbon of resonators over an infinite plate. The unit cell highlighted in red contains a domain wall between two phases with different valley Chern numbers. Top) band structure of the finite system, neglecting non-bulk modes, i. e. modes in the interior of the free dispersion curve $\Omega a=\left( \frac{k_x}{\pi}\right)^2$. Bottom) real space displacement field along the supercell sites for different frequencies and momenta (eigenvectors) as indicated with colored dots on the band structure. The two lines represent site A and site B.
 }
\label{grapheneDomainWall1}
\end{figure}

\begin{figure}
\epsfig{file=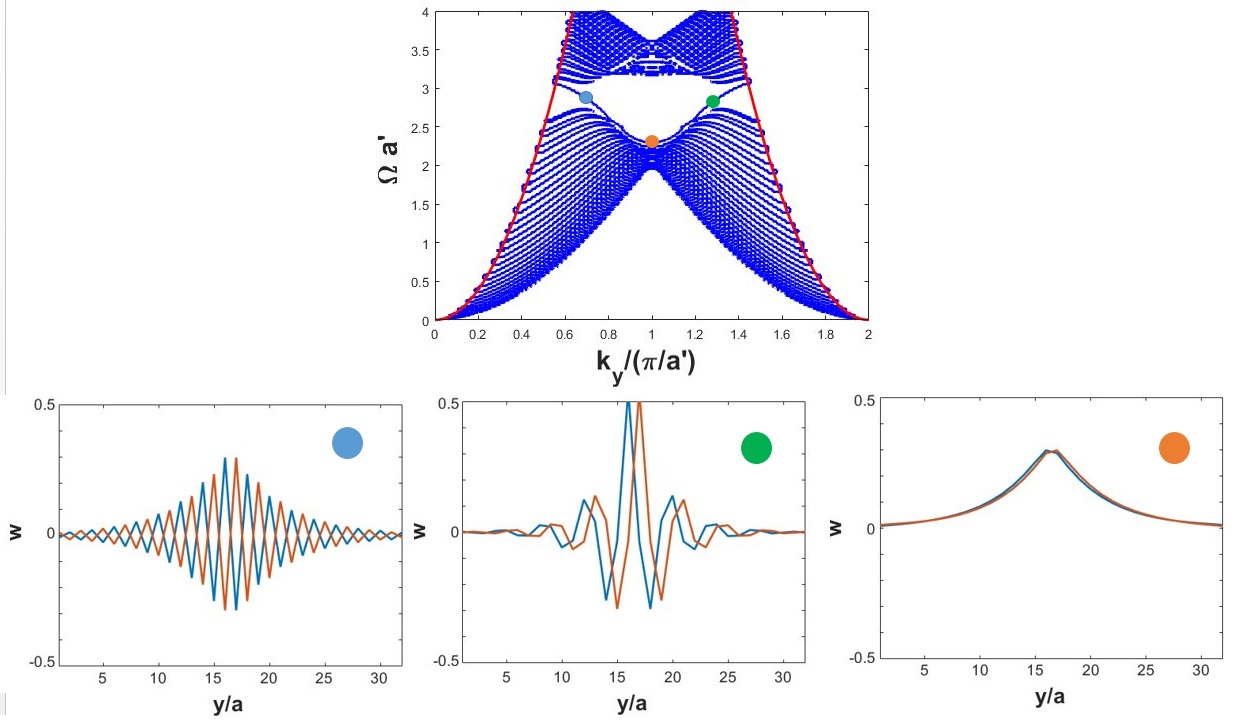,width=15cm}\\
\caption{(Color online)  Ribbon of resonators over an infinite plate. The unit cell highlighted in red contains a domain wall between two phases with different valley Chern numbers. Top) band structure of the finite system, neglecting non-bulk modes, i. e. modes in the interior of the free dispersion curve $\Omega a=\left( \frac{k_x}{\pi}\right)^2$. Bottom) real space displacement field along the supercell sites for different frequencies and momenta (eigenvectors) as indicated with colored dots on the band structure. The two lines represent site A and site B.
 }
\label{grapheneDomainWall2}
\end{figure}

As computed in Ref. \citep{Ruzzene_graphene}, the analogous to quantum valley Hall effect guarantees boundary modes localized at the interface between two phases. Inversion symmetry is broken by different masses of resonators in the two dimensional unit cell and two types of interface can be created (with zig-zag boundary). In this appendix we examine the symmetry of the boundary modes.
As explained in the main text, the ribbon structure has inversion symmetry at the domain wall provided the two phases are equally large and masses are the same. See Fig.\ref{graphenelattice}, 
full circles correspond to $\gamma=11$ and empty circles to $\gamma=9$ (the same at each side of the domain wall), all resonators have the same spring constant and their frequency is $\Omega_R=4\pi$. Dirac frequency for $\gamma=10$ is $\Omega_D=2.9$. \\
In section \ref{Sec:V}, ribbons such as the one shown in Fig. \ref{DomainWallT2_f15} contain different inversion symmetries at the domain wall, site 2 maps into itself from an inversion center different from where site 3 maps into site 1, and at the same time different from the inversion center where site 1 maps into site 3.\\
The symmetry of the boundary eigenvectors in presented in Fig. \ref{grapheneDomainWall1} and \ref{grapheneDomainWall2} for light and heavy boundary respectively. In the soft boundary, Fig. \ref{grapheneDomainWall1}, the mid gap band correspond to anti-symmetric modes. Symmetric boundary modes are lost in the bulk band structure. However, we can compute and plot the extended and symmetric boundary mode. In the hard boundary, Fig. \ref{grapheneDomainWall2}, the mid gap band merges with bulk bands near $k_x=\pi$. At each side, the symmetry is different, for $k_x<\pi$ modes are anti-symmetric under inversion symmetry and for $k_x>\pi$ modes are symmetric.

\end{document}